\documentclass[reprint,amsmath,amssymb,aps]{revtex4-2}
\usepackage{graphicx}
\usepackage{dcolumn}
\usepackage{gensymb}
\usepackage{hyperref}

\usepackage{adjustbox}
\usepackage{ulem}
\usepackage{cancel}

\usepackage{balance}
\usepackage{mathptmx}
\usepackage{graphicx} 
\usepackage{lastpage}
\usepackage[format=plain,justification=justified,singlelinecheck=false,font={stretch=1.125,small,sf},labelfont=bf,labelsep=space]{caption}
\usepackage{float}
\usepackage{fancyhdr}
\usepackage{fnpos}
\usepackage{booktabs}
\usepackage[english]{babel}

\usepackage{array}
\usepackage{droidsans}
\usepackage{charter}
\usepackage[T1]{fontenc}
\usepackage[usenames,dvipsnames]{xcolor}
\usepackage{setspace}
\usepackage[compact]{titlesec}
\usepackage{hyperref}
\renewcommand\labelenumi{(\roman{enumi})}
\renewcommand\theenumi\labelenumi

\usepackage{epstopdf}
\usepackage{ulem}
\usepackage{amssymb}

\definecolor{cream}{RGB}{222,217,201}

\usepackage[usenames,dvipsnames]{xcolor}
\newcommand{\Reinhard}[1]{\textcolor{orange}{#1}}

\newcommand{\Dominique}[1]{\textcolor{magenta}{#1}}

\begin{document}

\title{Coarsening transitions of wet liquid foams under microgravity conditions}%

\author{Marina Pasquet \textit{$^{a}$}, Nicolo Galvani\textit{$^{b,d}$}, Alice Requier\textit{$^{a}$},   Sylvie Cohen-Addad\textit{$^{b,c}$},  Reinhard Höhler \textit{$^{b,c}$},   Olivier Pitois \textit{$^{d}$},  Emmanuelle Rio \textit{$^{a}$},  Anniina Salonen \textit{$^{a}$} \& Dominique Langevin \textit{$^{a}$}}%

\affiliation{\\ }

\affiliation{$^{a}$Universit\'e Paris-Saclay, CNRS, Laboratoire de Physique des Solides, 91405, Orsay, France. }
\affiliation{$^{b}$~Sorbonne Universit\'e, CNRS-UMR 7588, Institut des NanoSciences de Paris, 4 place Jussieu, 75005 Paris, France.}
\affiliation{$^{c}$~Universit\'{e} Gustave Eiffel, 5 Bd Descartes, Champs-sur-Marne, F-77454 Marne-la-Vall\'{e}e cedex 2, France. }
\affiliation{$^{d}$~Universit\'{e} Gustave Eiffel, ENPC, CNRS, Laboratoire Navier, 5 Bd Descartes, Champs-sur-Marne, F-77454 Marne-la-Vall\'{e}e cedex 2, France. \\}

\date{\today}%

\begin{abstract}
We report foam coarsening studies which were performed in the International Space Station (ISS)  to suppress drainage due to gravity.  
Foams and bubbly liquids with controlled liquid fractions $\phi$  between 15 and 50\%  were investigated to study the transition between bubble growth laws previously reported near the dry limit $\phi \rightarrow 0$  and the dilute limit  $\phi \rightarrow 1$ (Ostwald ripening). We determined the coarsening rates; for the driest foams and the bubbly liquids, they are in close agreement with theoretical predictions. We observe a sharp cross-over between the respective laws at a  critical value $\phi^*$. At liquid fractions beyond this  transition, neighboring bubbles are no longer all in contact, like at a jamming transition. Remarkably $\phi^*$ is significantly larger than the random close packing volume fraction of the bubbles $\phi_{\text{rcp}}$ which was determined independently. We attribute the differences between $\phi^*$ and  $\phi_{\text{rcp}}$ to a weakly adhesive bubble interaction that we have studied in complementary ground-based experiments.
\end{abstract}

\maketitle


\section{Introduction}

Foams are concentrated dispersions of gas bubbles in a liquid or solid matrix 
~\cite{ weaire2001physics,cantat2013foams}.
Solid foams, obtained by solidifying liquid foams,  are  light-weight materials used for  thermal or acoustic insulation or as construction materials. Liquid foams also have   many applications, such as detergency, flotation, and oil recovery. Even when they are stabilized by surface-active agents adsorbed to the gas liquid interfaces, such as surfactants, liquid foams are generally short-lived. Understanding and controlling foam stability is therefore   mandatory for applications. 
\par
\begin{figure*}
\centering
\includegraphics[width=0.95\linewidth]{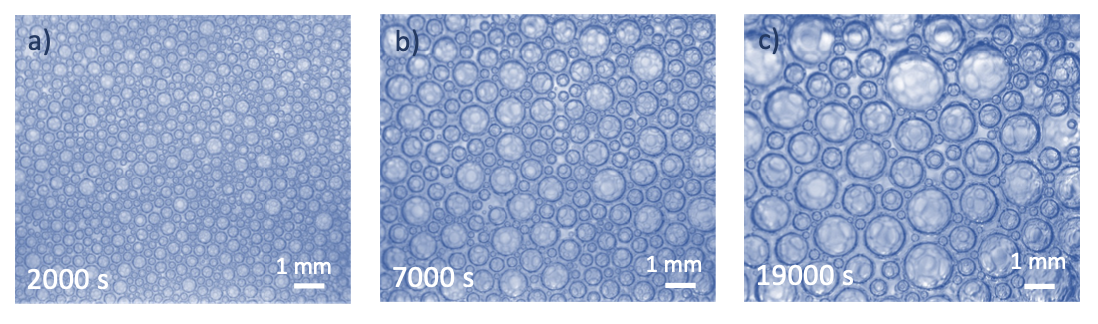}
\caption{
Foam coarsening in microgravity onboard the ISS. The liquid fraction is $\phi=25\%$. The times indicate the duration elapsed since the end of the sample foaming. }
\label{fig:structures}
\end{figure*}
\par
Foams are destabilized by three main processes: gravity induced drainage, coarsening of the structure due to the transfer of gas between bubbles driven by Laplace pressure differences, and bubble coalescence, due to the rupture of the liquid films between bubbles. These three processes are coupled in a complex manner~\cite{cantat2013foams}. For instance, coarsening and coalescence become faster when the liquid content of the foam decreases due to drainage. To study coarsening and coalescence under well controlled conditions, drainage needs to be suppressed. This can be achieved by rotating the sample in a clinostat, but the method is usually limited to foams with small liquid volume fractions, up to 15\%. 
A reliable way to study such foams is to perform experiments in microgravity. Even in the absence of drainage, it is still necessary to suppress one of the two remaining destabilisation processes, in order to study the other one accurately. To focus on coarsening, one can use efficient surfactants in appropriate concentrations to prevent coalescence ~\cite{cantat2013foams}. In this paper, we present a study of foam coarsening performed in the International Space Station (ISS) where gravity drainage is absent, so that we could study the coarsening of foams containing a wide range of  liquid volume fractions which are homogeneous throughout the sample and stable against coalescence.  

In  foams where the liquid volume fraction is small, called \textit{dry foams}, the bubbles are squeezed against each other so strongly that their shape is polyhedral. They are separated by liquid films, connected three by three through channels called \textit{Plateau borders}. With increasing liquid fraction $\phi$, the bubble shapes become more and more spherical. Beyond a critical liquid fraction $\phi $*, neighboring bubbles are no longer all in contact; such dispersions are called bubbly liquids. This jamming transition has a strong impact on the mechanical properties of foams and similar materials; it induces a transition from solid-like elastic to liquid-like viscous behavior~\cite{cantat2013foams}. For a random assembly of monodisperse hard spheres, $\phi $* is known to be about $36\%$ and corresponds to the random packing fraction $\phi_{\text{rcp}}$ of monodisperse spheres. This value decreases with increasing polydispersity of the spheres~\cite{Groot2009}. Jamming may be defined  either in terms of the appearance of bubble contact films or of the onset of solid-like mechanical behavior. In adhesive systems, these features are not necessarily correlated, in contrast to non-adhesive systems ~\cite{datta2011rheology}. 
In our work, we will use a definition based only on the existence of contact films, evidenced by the coarsening dynamics.\par
\par
The aim of our study is to investigate how coarsening is affected by the structural change near the jamming transition. In dry foams, gas is transferred among neighboring bubbles mainly by diffusion through the contact films, and there is a strong link between the topology of the packing and the concentration gradients driving gas exchange with neighbors. In the limit of large liquid fractions, bubbles form a dilute dispersion and exchange gas with the surrounding bulk liquid, which acts like a  reservoir of dissolved gas. This process is known as Ostwald ripening. In both cases, the bubble growth is governed by asymptotic power laws, but their exponents are different. The transition between these two limiting cases and its relation with the structural change at the jamming transition are not yet well understood and the experiments with foams and emulsions do not yet provide a clear global picture of coarsening. Note that grain growth in solid dispersions is similar to foam coarsening, but it is  more complex, due to the anisotropy of the crystal lattice ~\cite{baldan2002review}.

\par
We have therefore performed coarsening experiments using samples with liquid fractions in the range $15\% \leq \phi \leq 50\%$, covering both sides of the jamming transition. Figure~\ref{fig:structures} illustrates the coarsening of a foam with a liquid fraction of $25\%$ obtained in our experiments. 
The behavior of these dispersions not only depends on liquid fraction, it can also be modified by attractive interactions ~\cite{princen1983rheology, trappe2001}. Attraction can enhance contact forces and film areas at liquid fractions above $\phi_{\text{rcp}}$, and give rise to soft solid-like behavior for $\phi >\phi_{\text{rcp}}$ where bubbles form gel-like networks. The mechanical behavior of gelled droplet dispersions~\cite{fuhrmann2022, dong2022} has been investigated recently, but to our knowledge coarsening studies of gelled bubble or droplet suspensions  have not been reported. We work with weakly attractive bubbles. We have determined the contact angle that characterizes this attraction and discuss its relation to the observed coarsening behavior, which differs from predictions for repulsive bubble packings.  \par

Theoretical analyses and simulations of  coarsening in foams~\cite{thomas20153d}, emulsions ~\cite{taylor1998ostwald} and annealed metals~\cite{baldan2002review} predict, in agreement with experimental results, that after a transient regime, the distribution of bubble, droplet or grain radii, normalized by their average value,  becomes invariant with time. In this so-called scaling state, the particle growth is statistically self-similar. 

Growth models have been established for either small or large continuous phase volume fractions. In these models, the predicted average radius can be defined with
different weightings, but in the self-similar regime the different average radii are proportional to each other. In the limit of small liquid fractions $\phi \rightarrow 0$ (dry foam limit), the  average radius $\left< R\right>$ is predicted to increase with time $t$ following a parabolic law:
\begin{equation}
\left< {R}(t)\right>^2 = \left<{R_o}\right>^2+\Omega_{p}(t-t_o)
\label{eq:GrowthLawNeuman}
\end{equation}
The radius $R(t)$ is an equivalent radius, proportional to the square root of the area of the bubbles which are not spherical. In the opposite limit where $\phi \rightarrow 1$ (dilute bubbly liquid),  the average  radius is predicted to increase with time as: 
\begin{equation}
\left< {R}(t)\right>^3 = \left<{R_o}\right>^3+\Omega_{c}(t-t_o)
\label{eq:GrowthLawOstwald}
\end{equation}
The growth exponents do not depend on the choice of a specific average radius, but the growth rates $\Omega_{p}$ and $\Omega_{r}$ will depend it.
\par
The limit of small liquid fraction was first analyzed by Neuman~\cite{von1952metal}, Wagner~\cite{Wagner1961} and Mullins~\cite{mullins1986statistical}, and confirmed by experiments with dry foams~\cite{durian1991scaling, lambert2007experimental}.
The opposite limit, called Ostwald ripening, was studied by Lifshitz, Slyozov~\cite{Lifshitz1961} and Wagner~\cite{Wagner1961} (LSW), and confirmed by experiments with alloys~\cite{baldan2002review} and dilute emulsions~\cite{taylor1998ostwald}. In both equations $\left< {R_o}\right>$ is the average radius at a reference time $t_o$ in the scaling state.
 With increasing time, Eqs.~\ref{eq:GrowthLawNeuman}~and~\ref{eq:GrowthLawOstwald} respectively converge to  power laws: $\left< {R}\right> \sim t^{1/2}$ in dry foams and $\left< {R}\right> \sim t^{1/3}$ in dilute bubble dispersions. \par
The modification of the exponent as a function of liquid fraction is related to the mechanism of diffusive gas transfer between bubbles. 
In dry foams, it occurs mostly through the thin films separating a bubble from its neighbors, and whose thickness, of the  order of tens of nanometers, varies only weakly with bubble size. The flow is driven by the difference between the Laplace pressures in the bubble and in its neighbors.
In bubbly liquids, bubbles exchange gas with the surrounding liquid, whose gas concentration is  set by the average bubble pressure in the neighbors. However, in this case, the dissolved gas concentration decreases with the distance from the bubble surface over a range of the order of the bubble radius. The gas transfer in foams and bubbly liquids is thus set by two different length scales, in the first case the film thickness which is practically constant, and in the second case the bubble radius which evolves with time~\cite{mullins1986statistical}. Similar remarks apply to coarsening of emulsions.\par

The growth laws are thus well understood theoretically in the limiting cases of small and large liquid fractions but the knowledge at intermediate liquid fractions is still lacking.

In the following, after describing the experimental details, we will present 
our results on bubble growth laws and analyze the different regimes we have evidenced. 

\section{Materials and methods \label{sec:Methods}}

The experiments are performed aboard the International Space Station using the experiment container described in~\cite{born2021soft}. The residual gravity acceleration fluctuations on board the ISS are reported to be on the order of or less than $10^{-6} g$, for frequencies below 0.01 Hz \cite{NASAacceleration}. Each foam sample is placed in a hermetically closed transparent cell, containing the required volumes of foaming liquid and air to obtain a given liquid fraction $\phi$. The foam sample is generated \textit{in situ} using the back and forth actuation of the magnetic piston in the cell. The range of investigated liquid fractions lies between 15\% and 50\%. 
The foams are made using aqueous solutions of the surfactant Tetradecyl-Trimethyl-Ammonium Bromide TTAB (purchased from Sigma-Aldrich, purity $\ge 99$~\%) in ultrapure water. The TTAB concentration is $c=$~5 g/L, about 4 times larger than the critical micellar concentration (cmc = 1.2 g/L) so that the bubbles do not coalesce. The surface tension $\gamma$ of the TTAB solution measured at room temperature is: $\gamma= 37.1$~mN/m. \par

The experiment container is equipped with a camera that records images of the bubbles at the surface of the cell and with a probe based on multiple light scattering, Diffuse Transmission Spectroscopy (DTS) as described in~\cite{born2021soft}. The images are analyzed manually as described in~\cite{CRAS2023}. 
For each liquid fraction, we have measured the bubble size distributions as a function of foam age $t$, \textit{i.e.} time elapsed since the end of the foaming , and identified the Scaling State~\cite{PNAS}. Using these distributions, we calculated the different averages  $R_{ij}= \frac{R^i}{R^j}$. The number average radius is $R_{10}$ and the mean Sauter radius $R_{32}$. \par
 Diffuse Transmission Spectroscopy consists in measuring the light intensity diffusely transmitted through the sample. The average bubble radius in the bulk of foam is then deduced from the measured transmission coefficient for the specific geometry of the ISS set-up, as detailed in~\cite{CRAS2023}. The results showed that, for the same foam samples as studied here,  the evolution of the average bubble size observed at the surface and in the bulk are the same within experimental error~\cite{CRAS2023}. 
 We focus in this paper on the bubble growth laws $R_{32}(t) $ and will use the data from this earlier publication.
 \par
 Ground experiments were also performed to characterize the adhesion between bubbles. Monodisperse bubbles with the same TTAB concentration  as in the coarsening experiments were formed in a microfluidic T junction generator. The glassware, teflon tubing and the microfluidic device were carefully cleaned before use. The gas was air to which fluorinated vapour (perfluorohexane C$_6$F$_{14}$) was added in order to stop ripening and  to prevent the bubbles from becoming polydisperse.  
 Dilute bubble dispersions were injected into a hermetically closed observation cell,filled with the TTAB solution and covered by a glass plate without any air pocket. The contact angles were measured using image analysis and the method described in ~\cite{seknagi2022structure}.

\section{Experimental results}

\subsection{Bubble growth laws}
\label{sec:GrowthLaws}

\begin{figure}[h]
    \centering
     \includegraphics[width = 8.5cm]{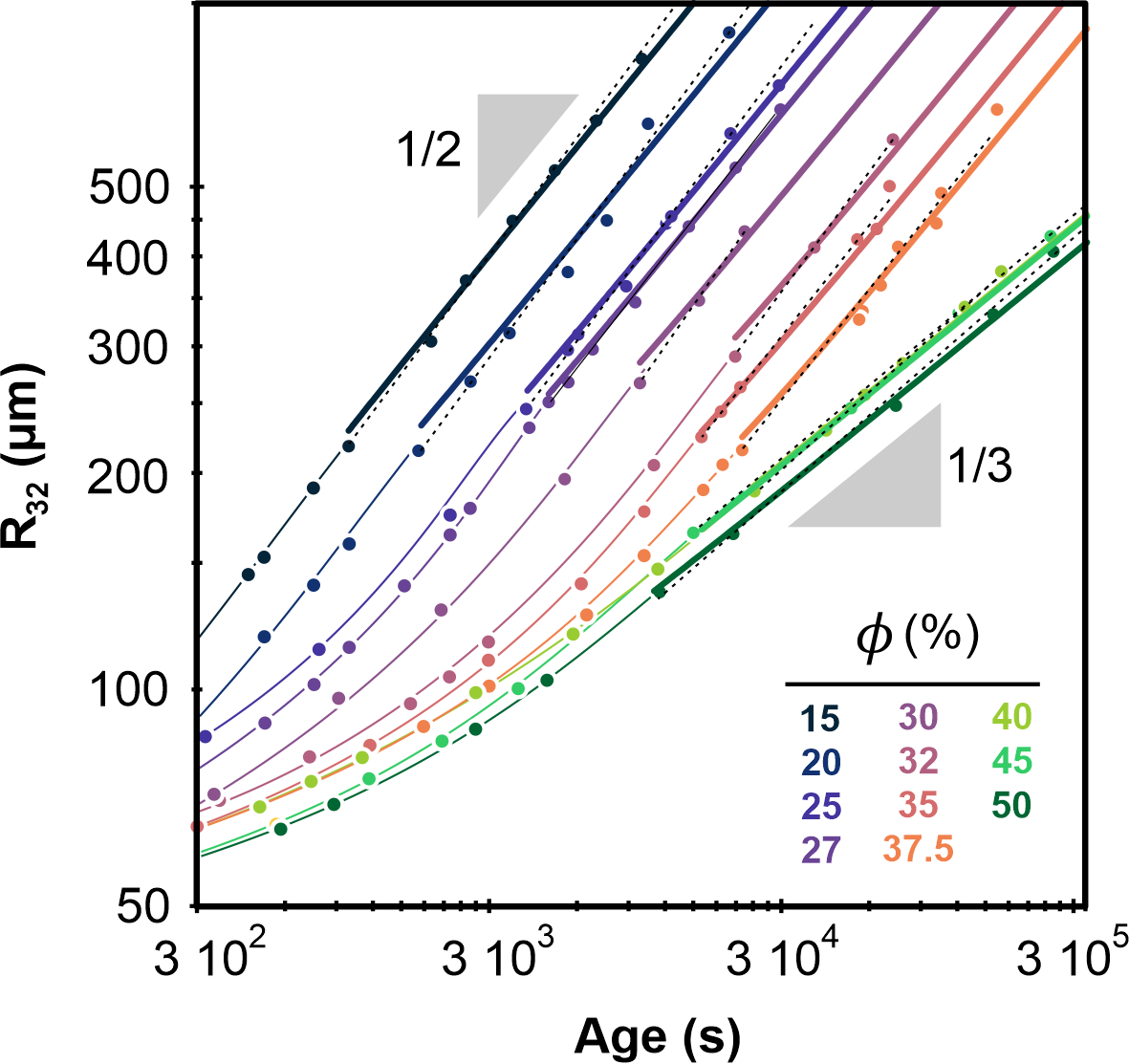}
    \caption{ Average bubble Sauter radius versus foam age for the different liquid fractions. The radii are measured with an accuracy of 15~$\mu$m. Dotted lines represent power laws $R_{32} \propto t^\alpha$ obtained by fits within the scaling state. Corresponding prefactors and exponents $\alpha$ are reported in table~\ref{tab:coarsening_rates}. The thick continuous lines represent growth laws within the scaling state, obtained by fitting a parabolic law to the data for $\phi \leq 37.5\%$ and a cubic law to the data for $\phi \geq 40\% $, as explained in the text. The fixed parameters $R_o =60\;\mu$m, and corresponding $t_o  $ are indicated in table~\ref{tab:coarsening_rates} for $\phi$ values increasing from 15\% to 50\% as labelled in the graph. The thin curves are extrapolations of the thick ones using equations 1 and 2 curves.}
    
    \label{fig:PowerLaw}
\end{figure}

Figure~\ref{fig:PowerLaw} shows the evolution of the Sauter mean bubble radius with foam age, deduced from image analysis and DTS light scattering data, for each investigated liquid fraction.
The bubble growth with age can be fitted in the Scaling State by  power laws: $R_{32} \sim t^{\alpha}$, with $\alpha \approx 1/2$ for $\phi \le 37.5\%$ and $\alpha \approx 1/3$ for $ \phi \ge 40\%$ (see table~\ref{tab:coarsening_rates}). The cross over between these two regimes occurs near a liquid fraction $\phi^{*}\approx 39\% $ and is rapid as illustrated in fig.~\ref{fig:exponents} of the Appendix.
\par

Our study of the bubble size distribution of the same coarsening foams~\cite{PNAS} shows that due to their polydispersity, the volume fraction that corresponds to the random close packing is : $\phi_{\text{rcp}} \approx 31\%$. Therefore, at $\phi > 31\%$ one would expect gas transfer to occur no longer through contact films but through bulk liquid, and the coarsening exponent should then be 1/3, in contradiction with our observations showing such an exponent only for $\phi > \phi^{*} > \phi_{\text{rcp}}$. Since the coarsening exponent 1/2 is a signature of a gas transfer mechanism predominantly through the contact films between the bubbles, our results suggest that the films persist far above $\phi_{\text{rcp}}$, and up to $\phi^{*}$. We attribute this effect to adhesive bubbles interactions that promote contact films as shown by observations reported in the section~\ref{sec:Attraction}.\par

By reminiscence of the dry foam and the dilute bubble dispersion respectively, we name the regimes we observed: 
\begin{enumerate}
    \item The adhesive foam regime, for $\phi < \phi^{*}$.
    \item The bubbly liquid regime, for $\phi > \phi^{*} $. \end{enumerate}
In the adhesive foam regime, we expect the growth rate to be described by~Eq.~\ref{eq:GrowthLawNeuman}.
We determine the coarsening rate $\Omega_{p}$ for each $\phi$, by fitting to the data the growth law:
 $ R_{32}^2(t) = R_o^2 + \Omega_{p,R_{32}} \; (t-t_o)$.
Since the foaming process yields samples whose initial bubble radius $R_{32}$ is close to 60~$\mu$m, we choose for all $\phi$ a reference radius $R_o =60\;\mu$m and the corresponding reference time $t_o$. 
From the bubble size distribution analysis~\cite{PNAS}, we obtain the time (and bubble size) where the foam has reached the scaling state. For all investigated liquid fractions, this happens for $R_{32} \gtrsim 250\;\mu$m. Thus we fit the parameter $\Omega_{p,R_{32}}$ in the corresponding range of times in the scaling state.  
We do a similar analysis to determine the coarsening rate $\Omega_{c,R_{32}}$ in the bubbly liquid regime with the growth law: 
$R_{32}^3(t) = R_o^3 + \Omega_{c,R_{32}} \; (t-t_o)$.
As can be seen in Figure~\ref{fig:PowerLaw}, all of these growth laws fit well to the data in each regime. The values of the coarsening rates are reported in Table ~\ref{tab:coarsening_rates}. 
We observe that the coarsening slows down as the liquid fraction increases in each regime. In sections~\ref{sec:DiscussionDryFoams} and~\ref{sec:DiscussionBubbly}, we discuss quantitatively the dependency of $\Omega_{p,R_{32}}$  and  $\Omega_{c,R_{32}}$ with liquid fraction.

\begin{table*}[]
    \centering
    \begin{tabular}{|c|c|c|c|c|c|c|c|c|c|c|c|}
    \toprule
        $\phi (\%)$ & 15 & 20 & 25 & 27 & 30 & 32 & 35 & 37.5 & 40 & 45 & 50 \\
    \midrule 
        $R_{32} / R_{21} $ & 1.3 & 1.3 & 1.3 & 1.3 & 1.3 & 1.2 & 1.2 & 1.2 & 1.1 & 1.1 & 1.1 \\
        $R_{32} / R_{10} $ & 2.0 & 2.2 & 2.0 & 2.0 & 1.9 & 1.7 & 1.6 & 1.6 & 1.2 & 1.2 & 1.1 \\
        Polydispersity & 0.38 & 0.41 & 0.37 & 0.37 & 0.34 & 0.27 & 0.25 & 0.24 & 0.09 & 0.09 & 0.06 \\
         $t_o$ (s) & 87& 78& 68& 68& 68& 68& 68& 284& 243& 243& 214 \\
    \midrule  
        Fitted exponent $\alpha$ & 0.55 & 0.55 & 0.55 & 0.52 & 0.59 & 0.55 & 0.52 & 0.51 & 0.36 & 0.35 & 0.35 \\
        Prefactor ($\mu$m/s$^{\alpha}$) & 5.00 & 3.48 & 2.60 & 3.17 & 1.20 & 1.18 & 1.51 & 1.36 & 4.90 & 5.79 & 4.90 \\
    \midrule  
        Exponent & 1/2 & 1/2 & 1/2 & 1/2 & 1/2 & 1/2 & 1/2 & 1/2 & 1/3 & 1/3 & 1/3 \\
        $\Omega_{p, R_{32}} $($\mu$m$^2$/s) & 55 & 31 & 16 & 13 & 7.9 & 4.3 & 3.2 & 2.1 & - & - & - \\
        $\Omega_{c, R_{32}} $($\mu$m$^3$/s) & - & - & - & - & - & - & - & - & 288 & 281 & 219 \\
    \bottomrule
    \end{tabular}
    \caption{Relations between average radii, coarsening rates and exponents evaluated in the scaling state. The polydispersity index is $p = R_{32}/{R_{30}}^{1/3}-1$\cite{Kraynik2004}, with $R_{30}$ the third moment of radius distribution. The time $t_o$ is the reference time at which $R_{32}= 60$~$\mu$m. The measurement uncertainty on the coarsening rate is about 10\%.}
    \label{tab:coarsening_rates}
\end{table*}

\begin{figure}[h]
    \centering
    \includegraphics[width = 8cm]{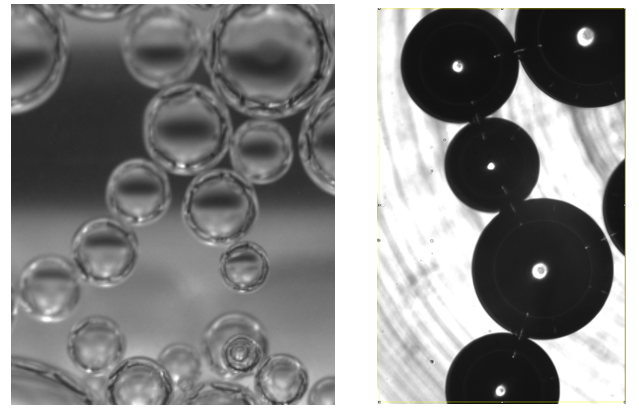}
    \caption{a) Bubble clusters surrounded by foaming liquid, spontaneously formed during our measurements in microgravity. b) Bubbles clusters confined under a horizontal transparent plate in the presence of gravity explained in section 4.2.  The black stripes are scratches in the bottom of the cell, which is far from the bubbles, the depth of the cell is 3~mm. 
    The bubble radii are of the order of $200~\mu$m in both a) and b). }
    \label{fig:adhesion}
\end{figure}

\subsection{Attraction between bubbles}
\label{sec:Attraction} 
In some microgravity experiments with large liquid fractions, we have observed regions with very few bubbles, organized in clusters. An example is illustrated in Figure \ref{fig:adhesion}a. Such a heterogeneous distribution of bubbles and the formation of aggregated clusters suggest the presence of an attractive bubble interaction. To verify the presence of attraction between bubbles dispersed in a TTAB solution, additional experiments were performed on ground.
Dilute dispersions of bubbles were confined under a transparent horizontal plate as described in section \ref{sec:Methods}. The image shown in Figure \ref{fig:adhesion}b confirms the spontaneous formation of clusters, indicating significant bubble adhesion in TTAB solutions. When the cell containing the bubble dispersion is tilted slightly, the bubbles migrate slowly along the transparent top plate, driven by buoyancy. This observation indicates that there are no pinning effects due to residual surface roughness of the top plate. The adhesive force  must be very small because  the bubbles are easily dispersed when the liquid in the cell is agitated. We observed similar adhesion with a smaller TTAB concentration, below the cmc (1 g/L), arguing against an effect due to micellar depletion which would disappear in this case. Similar clustering was also obtained at a larger TTAB concentration (20 g/L). These experiments were repeated with highly purified TTAB and the same results were obtained, arguing against impurity related artifacts. 

To characterize the adhesion quantitatively, the outline of two  bubbles of equal size in contact was observed in static equilibrium. By using an image analysis technique described in \cite{seknagi2022structure}, a contact angle of ($3.6 \pm 1^o$) was measured. This angle is larger than the one of  1$^{o}$  reported for Sodium Dodecyl Sulfate (another ionic surfactant) at similar concentrations~\cite{huisman1969contact,kralchevski1990formation}, perhaps due to the different molecular structures.

\section{Coarsening Models}
\label{sec:wetgrowth}

In this section, we present the existing predictions of the $\phi$-dependency of the coarsening rates $\Omega_{p}$ and $\Omega_{r}$. Then we extend theses predictions to account for the size of the contact films which is governed by the foam osmotic pressure, for the film curvature or for concurrent gas transfer through the bulk liquid which is expected to be dominant as bubbles are nearly in contact when $\phi \rightarrow \phi_{\text{rcp}}$. 

\subsection{Bubble average growth rate in bubbly liquids}

Later extensions of the LSW theory predict that the coarsening rate in dilute dispersions increases as the dispersion becomes more concentrated ~\cite{brailsford1979dependence, baldan2002review}. Indeed, as $\phi$ diminishes, the field of dissolved species around each particle is screened by the presence of the other particles. Concentration gradients are confined
over shorter distances, which increases the rate of mass exchange, thus the coarsening rate, which can be expressed as:
\begin{equation}
    \Omega_c= \Omega_r \;g(\phi).
    \label{eq:Omegac}
\end{equation}

The LSW mean field theory for coarsening of dilute dispersions (\textit{i.e.} in the limit $\phi \rightarrow 1$) predicts that the coarsening rate $\Omega_r$ for the number average radius $R_{10}$ (first moment of the size distribution) is:
\begin{equation}
\Omega_{r, R_{10}} = \frac{8}{9}\,\gamma\; D_m \;\textit{He} \; V_m
\label{eq:Ostwald}
\end{equation}
where $\gamma$ is the liquid-gas surface tension, $D_m$  the gas diffusion coefficient in the liquid, $He$ the Henry solubility coefficient of the gas in the liquid (expressed in mole m$^{-3}$ Pa$^{-1}$), and $V_m$ the gas molar volume. 
The dimensionless function $g(\phi)$ expresses the dependency of the coarsening rate with $\phi$.
By construction, $g(\phi)=1$ for $\phi=1$. 
Different predictions of $g(\phi)$ have been proposed, mainly in order to describe the coarsening of grains in annealing alloys. 
Measurements in alloys with continuous phase volume fractions between 40\% and 80\% showed that $g(\phi)$ varies by a factor of about 3 (see reference ~\cite{baldan2002review}). In these systems, each phase is anisotropic and sometimes heterogeneous, and the grain boundaries  do not have a unique energy, in contrast to the interfacial tension is foams.    These features bring complexity in predicting  realistic kinetic evolutions.   
Experiments with emulsions of variable continuous phase volume fraction are scarce. They, however, also suggest that $g(\phi)$ does not change rapidly with $\phi$ (it increases by less than a factor 2 between $\phi$=90\% and $\phi$= 98\% in ref~\cite{ariyaprakai2010influence}). Like emulsions, bubbly liquids are simpler model systems than alloys, because each phase and their interface are ideally homogeneous and isotropic.
\par

\subsection{Bubble average growth rate in  foams}
 \label{sec:drygrowth}

Von Neumann~\cite{von1952metal} and more recently Mc~Pherson and Srolovitz~\cite{srolovitz} have shown that in extremely dry 2D and 3D foams ($\phi \rightarrow 0$), the number of neighbors of a bubble determines the curvature of its interfaces and thus, the Laplace pressure differences that drive diffusive gas exchange. Experiments ~\cite{lambert2007experimental} and simulations ~\cite{thomas20153d} have confirmed that the size of 3D bubbles with 13-17 faces remains quasi-stationary in 3D foams, while bubbles with a smaller number of faces shrink, and those with a larger number of faces grow. 
\par
 As long as the liquid fraction remains smaller than a few percent, the foam structure can be represented as a dry polyhedral "skeleton" decorated by slender Plateau borders that contain most of the liquid and that are connected at nodes~\cite{weaire2001physics}.  In this range of $\phi$, interbubble gas transfer is dominated by diffusion of gas through the contact films. It is driven by the Laplace pressure differences between neighboring bubbles, which is related to the film curvature.
On this basis, and by averaging over suitable distributions of bubble geometries investigated using Surface Evolver simulations~\cite{hilgenfeldt2001dynamics}, Hilgenfeldt {\it et al} have predicted a parabolic growth law for the bubble radius $R_{10}$  of the form of Eq.\ref{eq:GrowthLawNeuman}, valid for small liquid fraction (dry foams) with a coarsening rate: 
\begin{equation}
    \Omega_{p, R_{10}}= C \;\Omega_0 \;\; f(\phi)
    \label{eq:Omega_pHilgenfeld}
\end{equation}
where $C$ is a constant defined in Eq. \ref{eq:C_H} , $f(\phi)$ is the fraction of the bubble area covered by films and is given at small $\phi$ by~\cite{hilgenfeldt2001dynamics} 
:  
\begin{equation}
f_{\text{dry}}(\phi) = (1-1.52 \sqrt{\phi}) ^2
\label{eq:oldf(phi)}
\end{equation}
\par
The coarsening constant $ \Omega_0$   depends on the physicochemical properties of the foaming liquid and is given by:
\begin{equation}
    \Omega_0= \frac{\gamma\;D_m\;He\;V_m}{h}
    \label{eq:Omega_0}
\end{equation}
where $h$ is the film thickness. The quantity  $D_m\textit{He}\, V_m P/h$, where $P$ is the ambient pressure, is the foam film permeability $\kappa$. In the case of very thin films, the film permeability to gas is controlled by two processes:  diffusion through the core of the film of thickness $h$ and from permeation through the monolayers of surfactant molecules adsorbed on each side of the film, described by the monolayer permeability $\kappa_s$~\cite{Princen1965}. In this case, the \Dominique{film} permeability becomes:
\begin{equation}
\kappa = \frac{ D_m He \, V_m P}{h +2D_{\text{ms}}/\kappa_s}
\label{eq:permeability}
\end{equation}
where $D_{\text{ms}}$ is the gas diffusion coefficient in the monolayers. 
Actually, in the general case, $\Omega_0$ is given by:
\begin{equation}
     \quad \Omega_0=\frac{\gamma\kappa}{ P}
  \label{eq:Omega_dry_with_permeability}
\end{equation}
The constant $C$ introduced in eq.~\ref{eq:Omega_pHilgenfeld} is a geometrical factor depending on the foam structure~\cite{hilgenfeldt2001dynamics} :
 \begin{equation}
C = \frac{2^{5/3}\delta_A}{3^{1/3}\pi^{2/3}\delta_V^{1/3}\beta}
\label{eq:C_H}
\end{equation}
 $\delta_A$ and $\delta_V$ relate the typical Plateau border length $L$ to the bubble surface area $A = \delta_A L^2$  and bubble volume $V = \delta_V L^3$. For random monodisperse foams, numerical simulations yield :  $\delta_V \approx 11.3$ and $\delta_A \approx 27$~\cite{hilgenfeldt2001dynamics}.
$L$ is related to the radius $R$ of a sphere having the same volume as an average bubble: $\delta_V L^3 = 4\pi R^3 /3$. The coefficient $\beta$ relates the border length $L$ to the effective mean curvature $H \approx 1/(\beta L)$ of the interfaces through which gas is transferred. Hilgenfeldt \textit{ et al} predicted $\beta \approx 10$ for dry foams~\cite{hilgenfeldt2001dynamics}. It follows that at small $\phi$, $C \approx 1.24$.
\par

A mean field approach going back to Wagner and Lemlich and reviewed in~\cite{STEVENSON2010374,Pitois2012} may be used to model coarsening in wet foams, in the regime where the interbubble gas transfer is still dominated by the diffusion of gas through the contact films. The volume of a bubble of radius $R$ evolves due to gas exchange \textit{via}  films whose areas are assumed to scale as $R^2$, driven by a Laplace pressure difference expressed as $(1/R_{21}-1/R)$, where $R_{21}$ is the average bubble radius, defined as the ratio of the second and the first moment of the bubble size distribution. This specific ratio which arises from mass conservation of the gas species, sets the critical radius for a bubble of radius $R$ to shrink or grow~\cite{Wagner1961}. This writes~\cite{Lemlich1978,Pitois2012}:
\begin{equation}
\frac{dR}{dt}= 2\; \Omega_0 \;\; f(\phi) \left[\frac{1}{R_{21}}-\frac{1}{R}\right] 
\end{equation}
where $\Omega_0$ is given by Eq.~\ref{eq:Omega_dry_with_permeability} and $f(\phi)$ is the fraction of the bubble area covered by films. In the asymptotic scaling state, it yields a parabolic growth law for mean radius, as $R_{21}^2(t)=R_o^2 +\Omega_{p, R_{21}} (t-t_o)$, with:
  \begin{equation}
  \Omega_{p, R_{21}}=\Omega_0 \;\; f(\phi)
  \label{eq:Omega_p}
 \end{equation}
 \par

Recent theoretical work derived $f(\phi)$ for arbitrary polydispersity and  for liquid fractions up to the jamming point~\cite{hohler2021capillary}. The  average contact area of a bubble normalized by the surface area of a sphere of the same volume as the bubble is predicted to be, for $0<\phi \le \phi_{\text{rcp}}$:
\cite{hohler2021capillary} 
\begin{equation}
    f(\phi) =\frac{\tilde{\Pi}}{\tilde{\Pi}+ 2(1-\phi)}.
    \label{eq:f(phi)}
\end{equation}
 $\tilde{\Pi}$ is the foam osmotic pressure, normalized by $ \gamma / R_{32}$, where $R_{32}$ is the Sauter mean radius, ratio of the third to the second moments of the size distribution.  
 The following empirical relation describes experimental and simulation data for disordered foams over the full range of foam liquid fractions~\cite{Maestro2013}:      
\begin{equation}
\tilde{\Pi} =\frac{k (\phi -\phi_{\text{rcp}})^2}{\sqrt{\phi} }
\label{eq:Pi}
\end{equation}
For a disordered assembly of monodisperse bubbles $\phi_{\text{rcp}}=0.36$, $k = 3.2$. In polydisperse foams, both $\phi_{\text{rcp}}$ and $k $ are modified, as will be discussed in the section \ref{sec:DiscussionWetFoams}.
Note that Eq.\ref{eq:f(phi)} 
is in good agreement with $f_{\text{dry}}(\phi)$ for dry foams, but in contrast to an extrapolation of  $f_{\text{dry}}(\phi)$, it vanishes at $\phi=\phi_{\text{rcp}}$ as expected. Indeed,  $f(\phi)$ being the fraction of the bubble surface covered by films it should vanish at $\phi=\phi_{\text{rcp}}$ in the absence of adhesion.
 \par

As pointed out by Mullins~\cite{mullins1986statistical}, the successful prediction of the parabolic form of the bubble growth law by mean field models relies on the scaling state where statistical geometric properties of the bubble packing are invariant in time.  
In the next section, we consider the vicinity of the jamming transition where the film area vanishes and gas transfer occurs between close-by nearly spherical bubbles.

 \subsection{Coarsening of foam near the jamming transition}\label{sec:kissingbubbles}
Near the jamming transition, at liquid fractions slightly larger than $\phi_{\text{rcp}}$, neighboring bubbles do not touch and do not form contact films. In the absence of gravity, they are therefore approximately spherical, but in contrast to the dilute limit considered in the LSW theory, the gap separating the surfaces of neighboring bubbles is much smaller than a typical bubble radius.  Accurate theories of coarsening in this regime are so far not available. However, the gas transfer between two bubbles that nearly touch has been investigated.  Schimming and Durian have derived an expression for the diffusive gas transfer between two neighboring spherical bubbles of radii $R_1$ and $R_2$ immersed in a liquid \cite{Schimming2017}. For two bubbles of similar size $R_1 \approx R_2 \approx R$ and a minimal distance separating the bubbles that we will take equal to $h$, assuming $R \gg h$,  the predicted gas volume flow rate is:
\begin{equation}
    Q_{\text{bulk}}\approx {D_m\text{He}V_m \Delta P} \pi R\,  \ln(R/h),
    \label{eq:Qbulk}
\end{equation}
where $\Delta P$ is the Laplace pressure difference between the two bubbles, due to their size difference.\par

At liquid fractions slightly smaller than $\phi_{\text{rcp}}$, the bubbles have on average 6 contacts of area $A_c$ and the gas volume flow rate through each film is: 
\begin{equation}
    Q_{\text{film}}=\frac{D_m\text{He}V_m \Delta P}{h} A_c =  \frac{D_m\text{He}V_m \Delta P}{6 h} 4\pi R^2 f(\phi).
     \label{eq:Qfilm1}
\end{equation}

\subsection{Requirements for coarsening models beyond the mean-field approximation}
\label{sec:DiscussionWetFoams}
The coarsening models discussed successfully predict the exponent of the bubble growth law in foams where gas transfer among bubbles is dominated by diffusion through contact films. They make simplifying assumptions about the foam structure that have the advantage of enabling analytical solutions, but which are too schematic to predict quantitatively the prefactor of the bubble growth law and its dependence on liquid fraction, especially in wet foams, as we will show in the following sections. We therefore revisit the derivation of the models for liquid fractions that are so large that the bubble shape is approximately spherical, but still small enough for diffusion through contact films to be the dominant mechanism of gas transfer among neighboring bubbles.  Lemlich's model  presented in section  ~\ref{sec:drygrowth}~\cite{Lemlich1978} is based on a mean field approximation, where every bubble exchanges gas with a fictive bubble of average size, while Hilgenfeldt {\it et al} ~\cite{hilgenfeldt2001dynamics} use empirical expressions for the bubble contact area and the Laplace pressure, derived from simulation for nearly dry foams: these expressions cannot be extrapolated to wet foams, whose structure is fundamentally different. To investigate how such models could be improved to obtain more quantitative predictions, we start from a representation of a wet foam as a packing of approximately spherical bubbles, as Lemlich does, but we do not make a mean field approximation and our discussion is not limited to the scaling state. We label bubbles by integer indices $i$ or $j$, ranging from 1 to the total number of bubbles in the foam that we call $N$.
The rate at which the volume of a gas bubble number $j$, of radius $R_j$  changes with time is in this case 
\begin{equation}
\frac{d}{dt} \frac{4\pi R_j^3}{3}=   \frac{4\pi R_j^2 2\gamma D_m\text{He}V_m}{h}  \sum_{i=1}^N f_{ij} \left(\frac{1}{R_j}-\frac{1}{R_i}\right)
\end{equation}   
$f_{ij}$ is the area fraction on the surface of  bubble $j$ covered by the contact with bubble number $i$; the sum is calculated over all of the neighboring bubbles where $f_i>0$. From this expression we deduce the time derivative of $R_j$ and we perform an arithmetic average  over all possible values of $j$, represented by angular brackets:
\begin{equation}
\frac{d}{dt} \left< R_j\right>=   
 2 \Omega_o  \left< \sum_{i=1}^N f_{ij}\left(\frac{1}{R_j}-\frac{1}{R_i}\right)\right>
 \label{eq:R2t}
\end{equation}
Eq.\ref{eq:R2t} shows that to make the prediction of Lemlich's mean field model more accurate, the arithmetic average of the contact area $f(\phi) = \left< f_{ij} \right>$ should be replaced by: 
\begin{equation}
 \left< \sum_{i=1}^N f_{ij}\left(\frac{1}{R_j}-\frac{1}{R_i}\right)\right>
/\left(\frac{1}{R_{21}}-\frac{1}{R}\right)
\end{equation}
\par 
To compare Eq.\ref{eq:R2t} with Hilgenfeldt's mean field model we write this equation as: 
\begin{equation}
\frac{d}{dt} \left< R_j\right>^2=   \frac{ 
 4 \gamma D_m\text{He}V_m }{\beta^* h} f(\phi)  
  \label{eq:omegawet}
\end{equation}
with 
\begin{equation}
\frac{1}{\beta^*}= \frac{\left<R_j\right>}{f(\phi)}\left< \sum_{i=1}^N f_{ij}\left(\frac{1}{R_j}-\frac{ 1}{R_i}\right)\right>
\end{equation}
From equations~\ref{eq:Omega_pHilgenfeld}, \ref{eq:Omega_0}, \ref{eq:C_H} and \ref{eq:omegawet}, we see that
in the wet case, the factor $1/\beta^*$ plays the role of the factor $1/\beta$ in the dry case. \par

To summarize, the coarsening process depends on the average Laplace pressure difference  with respect to the neighbors of each bubble (i.e. the contact film curvature), as well as on the respective contact areas. Our  analysis shows that to model foam coarsening quantitatively, these features must not be averaged independently, their correlation matters, and it can be expected to depend on liquid fraction and polydispersity.
 A deeper understanding of the packing geometry of wet coarsening foams is thus needed, simulations investigating this are under way.  
  Eq.  \ref{eq:omegawet} also shows that in the scaling state where all ratios of characteristic lengths of the structure are invariant in time, $\beta^*$ is independent of time, as expected.

\section{Discussion and comparison with models \label{sec:wetdiscussion}}
 
\subsection{Coarsening rate in the adhesive foam regime $\phi \leq \phi^{*}$}
\label{sec:DiscussionDryFoams}
Figure~\ref{fig:CoarseningRates} shows the decrease of the coarsening rate $\Omega_{p,R_{32}}$ as the liquid fraction  approaches $\phi^{*}$.
Because the coarsening exponent 1/2 is a signature of gas transfer  through contact films between bubbles, we conclude that such films persist above $\phi_{\text{rcp}}$ up to $\phi^{*}$. We attribute this to adhesive bubbles interactions that promote contact films evidenced by the observations reported in section~\ref{sec:Attraction}.
For $\phi \ll \phi_{\text{rcp}}$, the film area is expected to be set by the osmotic pressure predicted for non-adhesive foams by (Eq.\ref{eq:f(phi)}), because here, capillary contact forces dominate over adhesive forces. In contrast, for $\phi_{\text{rcp}} \leq \phi \leq \phi^{*}$, the film size is maintained at a non-zero value due to attractive interactions. We will discuss each of these two regimes in the following.
 
\subsubsection{Coarsening for $\phi \leq \phi_{\text{rcp}}$  }
\label{sec:DiscussionFoamRegime}
In this section we will first discuss how the area, thickness and permeability of contact films can be modelled as a function of liquid fraction, bubble size and physicochemical parameters, and then use this information to compare Lemlich's mean field model prediction  to our data.\par
If attractive bubble interactions are negligible, the fraction of bubble interfaces covered by contact films can be expressed as a function of the osmotic pressure $\tilde\Pi$ using Eq.~\ref{eq:f(phi)}. For monodisperse or weakly polydisperse foams  where the random close packing fraction  is $\phi_{\text{rcp}}\approx 36\%$  the empirical expression Eq.~\ref{eq:Pi} can be used to predict $\tilde\Pi$. However, for our strongly polydisperse samples where  $\phi_{\text{rcp}}$ is close to $ 31\%$,  this model has to be generalized. The change of $\phi_{\text{rcp}}$ implies a change of the parameter $k$ in Eq.~\ref{eq:Pi}. 
To explain this, we briefly revisit the derivation of Eq.~\ref{eq:Pi}. The factor $\phi^{-1/2}$ which dominates the variation of osmotic pressure with liquid fraction in the dry limit is set by the geometry of Plateau borders and the Laplace law~\cite{cantat2013foams}, the functional form of the factor $(\phi-\phi_{\text{rcp}})^2$ which dominates in the wet limit is derived empirically. The coefficient $k$ could simply be fitted to experimental osmotic pressure data, but this would make it impossible to predict the osmotic pressure for samples with polydispersities where such data have not been measured previously. If only $\phi_{\text{rcp}}$  is known for such a structure, $k$ can be estimated from a theoretical constraint which has been derived by Princen~\cite{Princen1986}. As recalled in detail in appendix B of reference~\cite{Maestro2013}, it is a direct consequence of the definition of osmotic pressure in a foam $ -\gamma \Pi dV_{liquid} = \gamma dA$, relating the work required to extract a liquid volume $dV_{liquid}$ from a foam to the resulting change of interfacial energy, for a fixed total volume of the bubbles. We get:
\begin{equation}
   \int_0^{\phi_{\text{rcp}}}\frac{\tilde{\Pi}(\phi)}{(1-\phi)^2}d\phi = 3\left(A(0)/A(\phi_{\text{rcp}})-1\right) 
\end{equation}
 $A(\phi_{\text{rcp}})$ and $A(0)$ are the total interfacial bubble areas, respectively at the jamming point and in the dry foam limit.  Their ratio is a measure of the average deviation from a spherical shape of the bubbles in the dry limit. Extensive Surface Evolver studies by Kraynik {\it et al}~\cite{Kraynik2004} have shown that $A(0)/A(\phi_{\text{rcp}})$  hardly varies for a wide variety of foam structures, and that it is  close to  $1.1$. This determines the value of $k$ as a function of $\phi_{\text{rcp}}$ as follows:
\begin{equation}
    k(\phi_{\text{rcp}})=\frac{0.3}{(3 -\phi_{\text{rcp}}) \sqrt{\phi_{\text{rcp}}}+(\phi_{\text{rcp}}-1)(3+\phi_{\text{rcp}}) arctanh(\sqrt{\phi_{\text{rcp}}})},
    \label{eq:k}
\end{equation}
in full agreement with values given in the literature~\cite{Maestro2013}, \textit{i.e.} $k=7.5$ for $\phi_{\text{rcp}}=26\%$ (the case of a fcc packing) and $k=3.2$ for $\phi_{\text{rcp}}=36\%$ (monodisperse random close packing). The variation of $k$ with $\phi_{\text{rcp}}$ is illustrated by Figure~\ref{fig:kphi}, in the appendix.
For $\phi_{\text{rcp}}=31\%$, the case relevant for our samples, Eq.~\ref{eq:k} predicts $k=4.75$.  \par
These results can be used to estimate contact areas for polydisperse foams in the absence of attractive bubble interactions (contact angle zero) using Eq.~\ref{eq:f(phi)}. As shown in the section~\ref{sec:Attraction}, there is evidence of attractive bubble interactions in our samples. As a consequence, the angle between interfaces meeting at a contact is increased, leading to an enhanced contact area. We expect this effect to become negligible in dry foams where contact areas are large and we will discuss the impact of this effect at the end of this section.
\par
\begin{figure}[h]
    \centering
    \includegraphics[width = 8cm]{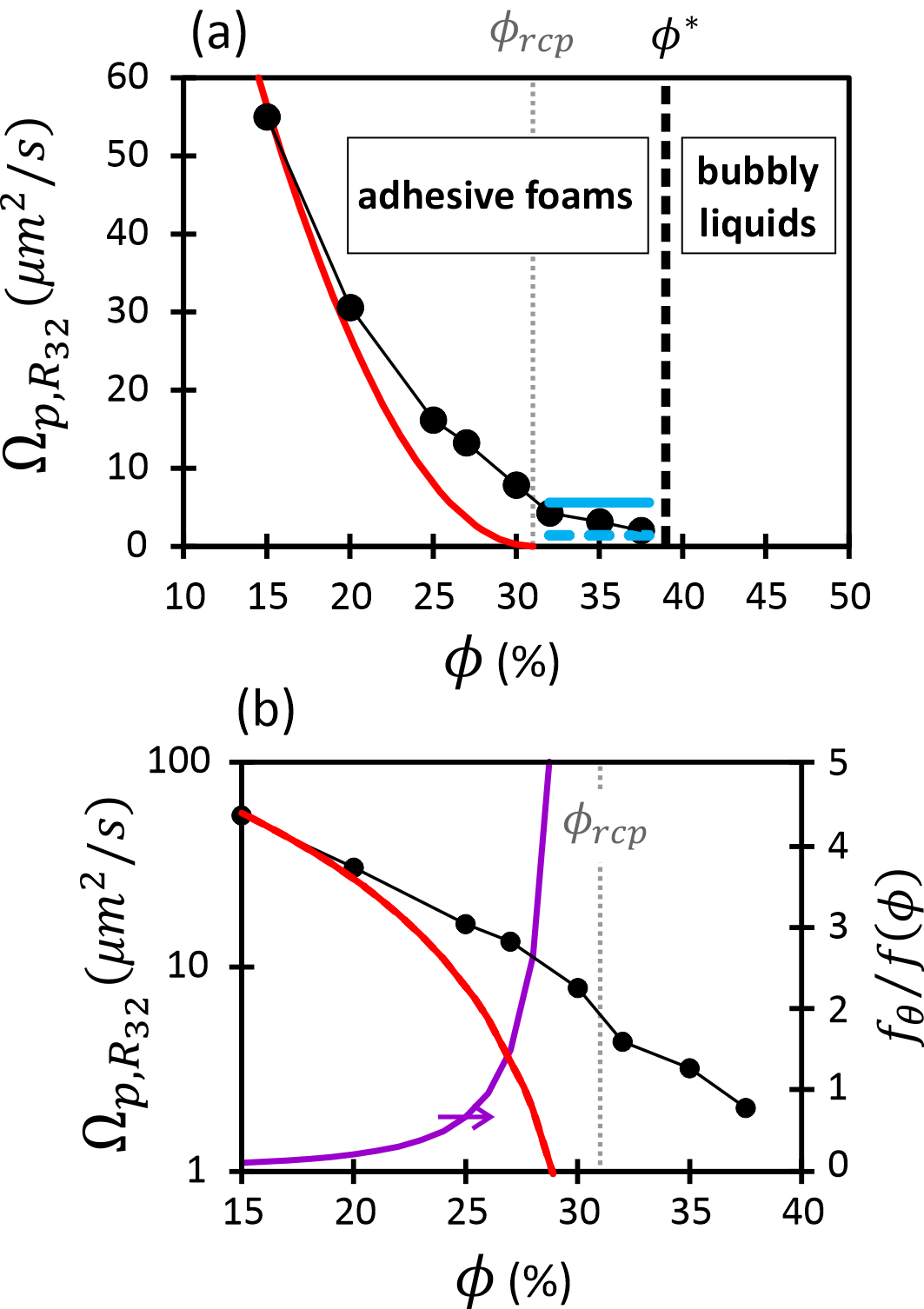}
    \caption{Coarsening rates for wet foams presented in linear scale (a) and logarithmic scale (b), as a function of liquid fraction. The plotted parameter is called $\Omega_{p,R_{32}}$ in the text.
    The red line represents the theoretical prediction, i.e. Eqs.~\ref{eq:Omega_p} and \ref{eq:f(phi)}, evaluated as explained in the text with $\phi_{\text{rcp}}=$31\%, $k = 4.75$. The horizontal blue lines correspond to the expected coarsening rates due solely to adhesion-induced contact films between bubbles (see equation~\ref{eq:ftheta}), assuming respectively 12 (continuous line) and 3 (dashed line) contacts per bubble. In (b), the purple curve represents the ratio of film areas induced solely by adhesion forces (Eq.~\ref{eq:ftheta}) and those induced solely by the osmotic pressure (Eq.~\ref{eq:f(phi)}).} 
    \label{fig:CoarseningRates}
\end{figure}

We have compared the measured coarsening rates for $\phi<\phi_{rcp}=31\%$ to that predicted by Lemlich model Eq.~\ref{eq:Omega_p}  with the film area given by Eq.~\ref{eq:f(phi)}, \ref{eq:Pi} and~\ref{eq:k}. For the comparison, the predicted $\Omega_{p, R_{21}}$ must be multiplied by the ratio $f(15\%)(R_{32}/R_{21})^2$. This scaling is justified since in the scaling state, the ratio $R_{32}/R_{21}$ does not depend on the liquid fraction as shown in table~1. 
We observe in Fig.~\ref{fig:CoarseningRates} that the decrease of $\Omega_{p,R32}$ with $\phi$ is much slower than predicted. This  is likely due to bubble adhesion which increases the area of contact films. This effect is most striking at $\phi_{\text{rcp}}$ where, in the absence of attractive interactions, the contact area should be zero, while a finite contact area will persist for positive contact angles as will be discussed in section~\ref{sec:DiscussionAdhesion}\par
Let us now compare the value of the observed growth rate to the rate predicted by Lemlich for $\phi=15\%$. This is the smallest investigated liquid fraction where the effect of adhesion is expected to be negligible. We get $\Omega_0=\Omega_{p,R32} \;(R_{21}/R_{32})^2 \;/ f(15\%)\approx 214 \pm 21~\mu$m$^2$s$^{-1}$. This value can be compared to that previously measured in quasi two-dimensional foams maintained at a constant capillary pressure (comparable to those in the current experiments)
 ~\cite{Forel2016}. In this work, the finite liquid fraction corrections proposed by Schimming and Durian \cite{Schimming2017} were used, leading to $\Omega_0 = 256$ $\mu$m$^2$.s$^{-1}$. This is in good agreement with the value measured in 3D foams, taking into account experimental uncertainties and the possible limitations of the models used. 
\par
Besides the film area, another major feature of the foam structure that determines coarsening rates is the average thickness of contact films. It results from an equilibrium between the film disjoining pressure $\Pi_{\text{disj}}$ and the capillary pressure $P_c$ that sets the difference between the liquid pressure in the Plateau borders and the gas pressure in the bubbles. 
For a foam, in the absence of adhesion or when it can be neglected as it is the case for $\phi<\phi_{\text{rcp}}$, the capillary pressure is given by~\cite{hohler2021capillary}:
\begin{equation}
    \frac{P_c}{\gamma / R_{32}}=\frac{\tilde{\Pi}}{1-\phi} +2
    \label{eq:Pc}
\end{equation}
In the range $15\% \leq \phi \leq 30\%$,  for average bubble sizes in the range $60$ $\mu$m$ < R_{32} < 800$ $\mu$m, the capillary pressure typically varies between 100 Pa and 1500 Pa.\par
Previous measurements of the disjoining pressure of films made with TTAB solutions at the cmc  showed that they are common black films and that the disjoining pressure is well predicted by DLVO theory~\cite{bergeron1997disjoining}: $\Pi_{\text{disj}}$ is the sum of a non-retarded van der Waals attraction (air/water/air Hamaker constant 3.7 10$^{-20}$\;J) and of an electrostatic repulsion with a surface potential $\Psi_o = 130\;$mV. We assume that these findings apply to our foams when $\phi<\phi_{\text{rcp}}$.
The Debye electrostatic screening length $\lambda_D$ is a function of the solution ionic strength. In our experiments, the concentration is about four times larger than the cmc. Therefore, the ionic strength results from the contribution of the surfactant species up to the cmc and a contribution of the fraction $\alpha$ of free counterions that are not bound to the micelles. 
The following expression has been proposed for a 1:1 electrolyte~\cite{pashley1987double}:
\begin{equation}
    \lambda_D^{-2} = \frac{e^2}{\epsilon_r \;\epsilon_o \;k_B T}  (2 \;\text{cmc} + \alpha\; (c-\text{cmc}))
   \label{eq:Debye_length}
\end{equation}
where $e$ is the electron charge, $\epsilon_o$ the vacuum dielectric permittivity, $\epsilon_r = 78.5$ the relative water permittivity, $k_B $ the Boltzmann constant, and $T$ the absolute temperature. The fraction of free counterions in a TTAB micellar solution is~\cite{perger2007thermodynamics}: $\alpha \approx 0.25$. Using Eq.~\ref{eq:Debye_length}, we estimate the Debye length of our foaming solutions as: $\lambda_D=4.3$\;nm. \par
By solving the equation $\Pi_{\text{disj}}=P_c$, we  predict the equilibrium film thickness $h_{eq}$ as a function of $R_{32}$ and $\phi$. For a given $\phi$, as the bubble radius $R_{32}$ increases during coarsening, the capillary pressure will decrease and therefore the film thickness is expected to increase since $\Pi_{\text{disj}}$ decreases with film thickness. The analysis shows that, for each liquid fraction, the film thickness slightly increases with $R_{32}$ as $h_{eq}\sim R_{32}^{0.15}$. 
In view of our measurement accuracy (cf.~Fig.~\ref{fig:PowerLaw}), this small dependency is too small to be resolved experimentally. Moreover, the relative increase of $h_{eq}$ with liquid fraction varying between 15\% and 30\% is of the order of 2\%. We conclude that in the range of investigated liquid fractions and bubble radii corresponding to Fig.~\ref{fig:PowerLaw}, the film thickness can be considered to a good approximation as constant. Our calculation for the equilibrium film thickness gives $h_{eq} \approx 35 \pm 3$~nm.
\par
We now discuss the gas permeability of the films. To estimate it for air, which is a mixture of nitrogen and oxygen, we recall that the inverse of this permeability is the sum of the inverse permeabilities of the two gases weighted by their relative proportions~\cite{Princen1965,weaire2001physics}. Taking the  values of $\text{He}$ and $D_m$ for \Reinhard{di-}nitrogen and di-oxygen gases given in the literature ~\cite{sander2015atmos,ferrell1967diffusion}, we obtain an effective value for $D_m \text{He}$ of $1.46 \times 10^{-14}$ mol\;m$^{-1}$\;s$^{-1}$\;Pa$^{-1}$. Neglecting any resistance to gas transfer from the adsorbed monolayers, using Eq.~\ref{eq:Omega_0} and $h_{eq}=35$~nm, we predict $\Omega_0 \approx 390~\mu$m$^2$s$^{-1}$. 
This is somewhat larger than the values deduced from the experiments (i.e. 214$~\mu$m$^2$s$^{-1}$). Note that we did not consider any permeability reduction due to the interfacial layers.

Let us now discuss a feature that may be missing in previous coarsening models, besides those pointed out in section \ref{sec:DiscussionWetFoams}. 
The film thicknesses in foams could be larger than the equilibrium thicknesses, because during bubble rearrangements, some films disappear and others are created, initially with a thickness much larger than the equilibrium value; this effect will slow down coarsening if the time it takes to thin down a new film to the equilibrium thickness is comparable or longer than the average time between successive rearrangements for a given bubble in the foam.  A slowing down of foam coarsening has indeed been reported recently in foams under shear, above a shear rate corresponding to a time interval between shear induced rearrangements~\cite{saint2023foam}. 
Analyses of multiple light scattering  data acquired during the coarsening experiments on the ISS are under way to measure the rate of rearrangements, and to determine whether they can affect the coarsening rates.

\subsubsection{Coarsening for  $\phi_{\text{rcp}} \leq \phi \leq \phi^{*}$}
\label{sec:DiscussionAdhesion}

In this section we will show that adhesive bubble interactions can have a significant impact on the coarsening of wet foams near the jamming transition and in the bubbly liquid regime.\par

At the liquid fraction     $\phi_{\text{rcp}}$,  bubbles in a foam without adhesive interactions  no longer exert forces on each other, the areas of the contact films between neighbors shrink to zero. Gas can be transferred among neighboring bubbles only through the bulk liquid, over distances typically much larger than a contact film thickness. Therefore, the coarsening process is expected to slow down significantly, compared to dryer foams.\par
\begin{figure}[h]
    \centering
    \includegraphics[width = 8cm]{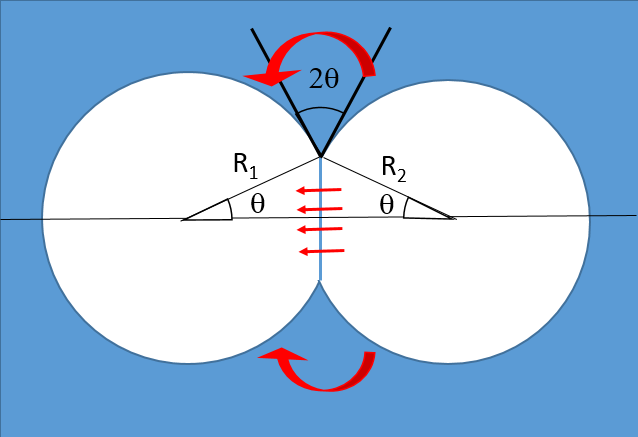}
    \caption{Side view of two adhesive bubbles, of radii $R_2 < R_1$, both very close to their average $R$. 
    In this schematic illustration, the contact angle $\theta$ is much larger than the value that we determined experimentally and that we consider in our model, and the radii are too close to be distinguished. The red arrows illustrate diffusive gas flow driven by the Laplace pressure difference, which can occur through the contact film as well as through the bulk liquid surrounding it.}
    \label{fig:cluster}
\end{figure}

In foams where interactions are adhesive, at liquid fractions $\phi \geq \phi_{\text{rcp}}$, neighboring bubbles spontaneously form contact films of typical area $A_c$. The energy cost of deforming the bubbles and increasing their surface area is compensated by the gain  $A_c F$ where $F$ is the  energy of adhesion per unit surface. $F$ also determines the contact angle $\theta$ at which the interfaces of neighboring bubbles meet, via the Young Dupré relation $F= 2\gamma (\cos\theta -1)$. Figure \ref{fig:cluster} illustrates this in the case of two adhesive 3D bubbles in contact. We consider this configuration as a minimal model of a bubble cluster, with two bubbles of slightly different radii $R_1$ and $R_2$, both close to their average $R$.
A simple geometrical calculation shows that for  this system, the contact area is $A_c= \pi R^2 sin^2\theta$.  \par
We will first discuss whether in such a bubble cluster the gas transfer among the bubbles is dominated by gas flow through the surrounding bulk liquid or through the contact film. 
\par
Equation~\ref{eq:Qfilm1} estimates the gas flow through contact films as a function of the difference of capillary pressures between two bubbles $\Delta P = 2\gamma (1/R_1 - 1/R_2)$  and under steady state conditions. Here, using $A_c= \pi R^2 sin^2\theta$, equation~\ref{eq:Qfilm1} becomes:
\begin{equation}
    Q_{\text{film}}=\frac{  D_m\text{He}V_m \Delta P  \,\pi R^2\, sin^2(\theta)}{ h} .
     \label{eq:Qfilm2}
\end{equation}

In the case of repulsive interactions where the  contact angle is $\theta=0$ and thus $A_c=0$, the gas transfer between two spherical neighboring bubbles that almost touch, with a minimal separation $h$, must occur through the bulk, in the meniscus surrounding the region where bubbles come close to each other. The  flow rate $ Q_{bulk}$ in this case is given by equation~\ref{eq:Qbulk}

\begin{figure}[h]
    \centering
    \includegraphics[width = 9cm]{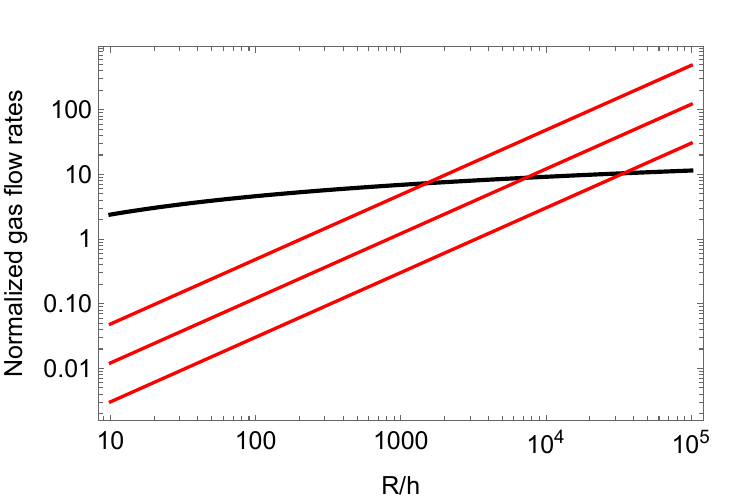}
    \caption
    {The gas transfer among two neighboring bubbles of similar sizes is plotted assuming either transfer going through a contact film (red lines, eq. \ref{eq:Qfilm2}) or through the bulk liquid surrounding the meniscus (black line, eq. \ref{eq:Qbulk}). The contact angles assumed in the cohesive case are, from top to bottom, 4, 2 and 1°. All the flow rates are normalized by $\pi R D_m\text{He}V_m\Delta P$.}
    \label{fig:ratio}
\end{figure}

Figure~\ref{fig:ratio} compares $ Q_{\text{film}}$ to $Q_{\text{bulk}}$, for several contact angles, near the one we have measured experimentally.  The figure illustrates that for any given contact angle of the order of a few degrees, in the limit of large ratios between the bubble radius and the liquid gap $h$ between neighboring bubbles, the gas transfer through the contact film will always become larger than the gas transfer through the bulk one would expect for repulsive bubbles with the same gap.  Indeed, the ratio of the two transfer rates scales as $(R/h)/\ln (R/h)$.  For a contact angle close to 4°, gas transfer through the films becomes dominant for  bubble radii more than $\approx 1500$ times the film thickness. For an equilibrium film thickness of the order of 40 nm, the critical bubble radius is thus of the order of $60$ $\mu$m. 
Figure 6 suggests that in our coarsening experiments where $R > 60~\mu$m, the gas transfer is dominated by diffusive flow through films even at liquid fractions larger than $\phi_{rcp}$, in agreement with the observed coarsening exponent close to 1/2.\par
Next, to estimate the coarsening rates, we recall that the function $f(\phi)$ in Equation~\ref{eq:f(phi)} represents the fraction of the bubble area covered by liquid films and vanishes for $\phi \geq \phi_{\text{rcp}}$. Here,  we assume weak adhesion forces between contacting bubbles, so that $f(\phi) = f_{\theta}$ for $\phi \geq \phi_{\text{rcp}}$. We estimate $f_{\theta}$ for $\phi \geq \phi_{\text{rcp}}$ using the following relation: 
\begin{equation}
     f_{\theta}=\frac{zA_c}{4\pi R^2}=\frac{z\, sin^2(\theta)}{4}
    \label{eq:ftheta}
\end{equation}
where $z$ is the average number of contacts per bubble. The coarsening rates obtained by equation \ref{eq:Omega_p} where we replace $f(\phi)$ by $f_{\theta}$ are plotted in blue in Figure~\ref{fig:CoarseningRates} for $z=12$ and $z=3$, showing agreement with the measured values in that range of liquid volume fraction. In non-adhesive foams $z$ would be equal to 6 for $\phi\approx \phi_{\text{rcp}}$. The number of neighbors in our adhesive foam samples cannot be deduced from our observations. The values of $z$ used in Figure~\ref{fig:CoarseningRates} illustrate the impact of this parameter in order of magnitude. We note that for $z>6$, the bubble network could support static stresses, so that Eq.\ref{eq:ftheta} where this effect is ignored only provides a rough approximation.  \\

The coarsening of adhesive foams raises many challenging theoretical questions whose solution is beyond the scope of the present paper. Coarsening generally amplifies the difference between the radii of neighboring bubbles. The contact film separating bubbles of different sizes is therefore increasingly curved. Moreover, adhesion reduces the interfacial tension in the contact film, which enhances the curvature. Therefore, for a given perimeter, the curvature of the contact film  enhances its area and thus the efficiency of gas transfer through it. To predict the evolution of bubble clusters or foams due to coarsening, the geometry of contact films needs to be investigated in detail. 
In addition, a  detailed investigation of the gas transfer through the bulk liquid near the contact line would be of interest, depending on the contact angle. 
Yet another important question concerns the connectivity of the flocculated bubble clusters that are expected at liquid fractions larger than $\phi^{*}$. The network of contacts must evolve as the gas in smaller bubbles is transferred to larger bubbles. Depending on liquid fraction, the contact network might percolate so that all bubbles are connected to each other, or it might split up into independent clusters. In each of these clusters, all the gas will ultimately be accumulated in a single remaining bubble, and if these bubbles remain unconnected, there would no longer be any effect of adhesion in the limit of a long coarsening time. Bubbles in clusters with multiple adhesive contacts may have a distorted shape that can no longer be modeled as a sphere. This feature and its effect on coarsening could be studied using simulations based on the Surface Evolver software.\par
In contrast to our results, a progressive transition of the coarsening law exponent from 1/2 to 1/3 over a liquid fraction interval   $23\% <  \phi < 38\%$  has been reported by Isert {\it et al} in a  foam coarsening experiment, where drainage was suppressed by magnetic levitation \cite{isert2013coarsening}. The transition onset  was  thus observed at much lower $\phi$, compared to our results. This could partly be due to a systematic underestimation of $\phi$, mentioned by the authors \cite{isert2015}, shifting the onset of the transition to $25\%$. The discrepancies  
could  also be due to the use of a highly concentrated SDS surfactant solution,  for which the contact angle could be significantly smaller than for TTAB. 
A  different analysis of our  data and of those of Isert {\it et al} has been proposed by Durian ~\cite{Durian2023} to explain the change of the coarsening exponent with liquid fraction, based on an ad hoc expression of the contact film radius dependence on liquid fraction.
\par
To assess if foam  coarsening  modified by adhesion is observable on Earth, we consider a  foam column floating on a liquid reservoir. 
At the top of the column, the cumulated buoyancy forces of the bubbles below tend to dominate over adhesive forces. At the bottom, the buoyancy force pushing bubbles against their neighbours above is $\rho g 4\pi R^3/3$; the contact area due to this force is estimated by dividing it by the capillary pressure $2\gamma/R$. This may be compared to the contact area solely due to adhesion in the absence of gravity  $A_c =\pi R^2 \sin^2(\theta) \approx \pi R^2 \theta ^2$ for $\theta \ll 1$ considered here (cf. Fig.~\ref{fig:cluster}). 
The contact areas induced by either gravity or adhesion are equal if $\theta \approx \sqrt{2\; \text{Bo}/3}$, where the Bond number is  $ \text{Bo}=\rho g R^2/\gamma$. For $\gamma = 37.1$ mN/m and a typical bubble radius $ R = 100\; \mu$m, we have $\theta\approx 3^{o}$ on Earth and $\theta\approx 0.003^{o}$ in the ISS. 
Therefore, the effect of  a small contact angle of $4^{o}$ characteristic of our foaming solution cannot be easily detected in typical 3D bulk foam coarsening experiments on Earth  whereas it is revealed under microgravity. Moreover, the adhesive networks for liquid fractions $\phi_{rcp} < \phi < \phi^*$ could be difficult to observe on Earth. Gravity induced creaming may crush these structures until their liquid fraction approaches $\phi_{rcp}$. 
 
\subsection{Coarsening rates in bubbly liquids $\phi>\phi^{*}$}
\label{sec:DiscussionBubbly}

At liquid fractions beyond the dilute limit, the average distance between bubbles decreases. Therefore, concentration gradients of the gas dissolved in the continuous phase do not extend independently up to an infinite distance around each bubble as assumed in the LSW theory, but over a shorter distance. As a consequence, the coarsening rate is expected to increase as $\phi$ decreases. Such an effect has been observed in annealing alloys and in emulsions, as discussed in the section~\ref{sec:wetgrowth}.
\par
To compare the measured coarsening rates $\Omega_{c,R_{32}}$ to those predicted for the mean radius growth law of concentrated particle dispersions, we must take into account the polydispersity which, in this regime, evolves with the liquid fraction. From measurements of the bubble size distributions~\cite{PNAS}, we determined the geometrical coefficient $C_o= (R_{32}/ R_{10})$ for each $\phi$ (see Table~\ref{tab:coarsening_rates}). We then calculated the coarsening rate $\Omega_{c, R_{10}}= \Omega_{c,R_{32}} / C_o^3$. 
These rates are much larger than the rate expected in the dilute limit (Eq.~\ref{eq:Ostwald}): $\Omega_{r} =12.1$~$\mu$m$^3.$s$^{-1}$. \par

\begin{figure}
    \centering
    \includegraphics[width = 8.5cm]{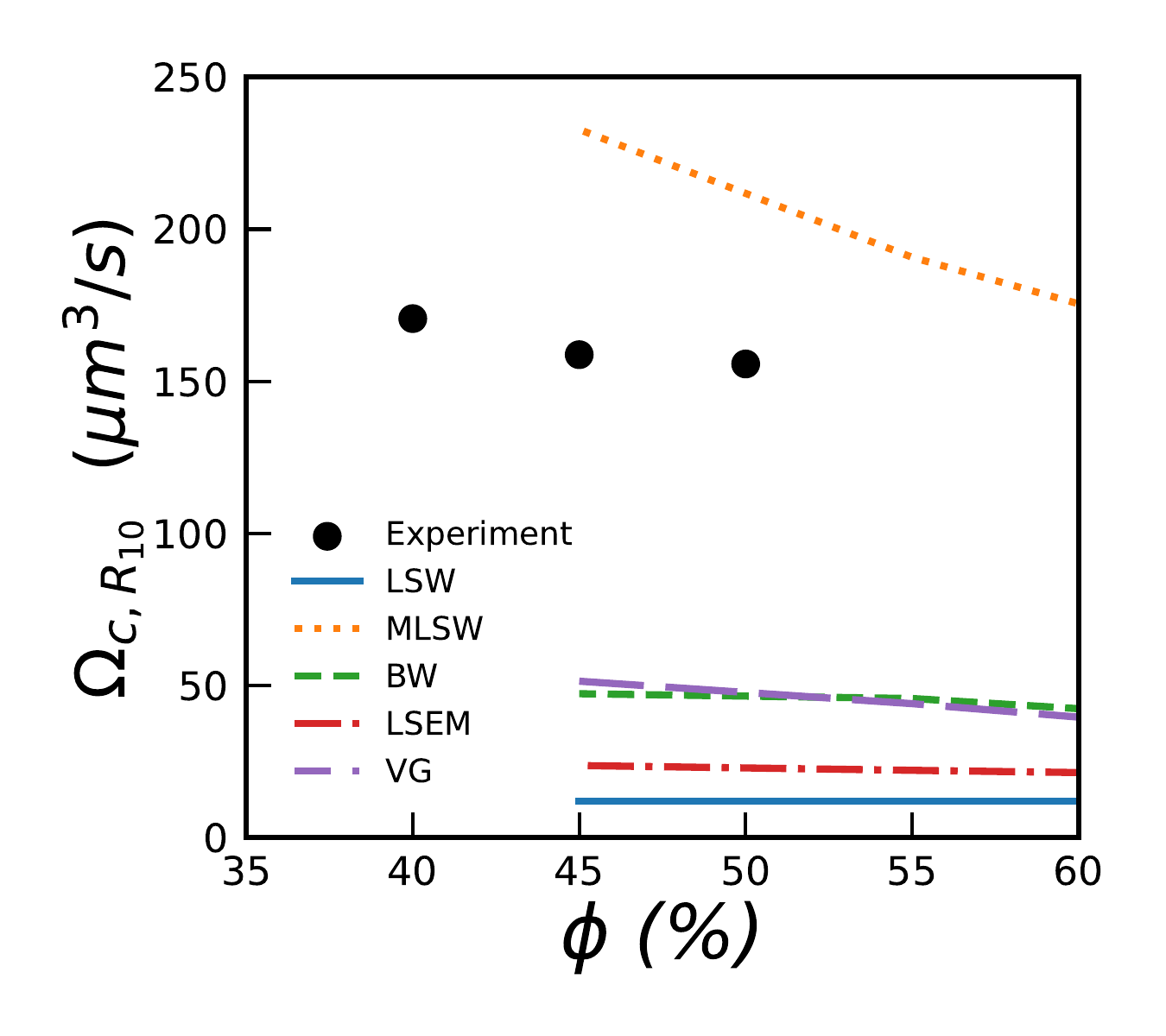}
    \caption{Coarsening rate $\Omega_{c,R_{10}}$  for bubbly liquids versus liquid fraction. 
    The lines show numerical predictions yielded by 5 theories presented in reviews\cite{Mahalingam1987, baldan2002review}. They  provide predictions only down to $45\%$.}
    \label{fig:Omegac_comparison}
\end{figure}

In Figure~\ref{fig:Omegac_comparison}, we show the measured values for $\Omega_{c,R_{10}}$ together with predictions from various simulations and theories~\cite{baldan2002review}. We see that most of them predict coarsening rates much smaller than the measured ones, except for the MLSW model.  \par

Experiments with emulsions showed that their coarsening rate was extremely dependent on the surfactant. Variations by several orders of magnitude for a given $\phi$, have been reported~\cite{roger2015emulsion} and attributed to exchanges between drops during collisions.  

\section{Conclusion}
Our experiments under microgravity show that during wet foam coarsening the average bubble radius increases with time as $t^{1/2}$ as in dry foams. The transition towards the Ostwald ripening regime (average bubble growth as $t^{1/3}$) is rather sharp and occurs at a liquid fraction close to $\phi^* =39\%$, higher than expected for the jamming transition of monodisperse hard spheres, $\phi^* \approx 36\% $ and far above the random close packing fraction $\phi_{\text{rcp}}\approx 31\%$ of spheres packing with the polydispersity of coarsening wet foams in the scaling state. We  deduced consistently the value of $\phi_{\text{rcp}}$ from simulations of polydisperse sphere packings ~\cite{PNAS} and from a theoretical a previous theoretical model, both linking $\phi_{rcp}$  to the experimentally observed polydispersity ~\cite{Groot2009}.

At liquid fractions below $25\%$, the observed coarsening rates agree approximately with previous models, the differences may be due to the contribution of the surfactant monolayers to film permeability, to non equilibrium films with enhanced thickness or  to the  mean field approximations used in the models.
 At liquid fractions larger than $25\%$ and in particular near $\phi_{\text{rcp}}$ the coarsening rates are much larger than predicted by 
 coarsening models where only gas transfer through films is considered. This  cannot be due to gas transfer through bulk liquid whose importance is expected to increase with $\phi$, because such a contribution  
 would be inconsistent with the coarsening exponent of 1/2 that we observe up to  $39\%$ and which is characteristic of gas transfer through films. We attribute these features at least partially to a weak attractive interaction between the bubbles, which we evidenced in dedicated ground based experiments. Adhesion is expected to enhance contact film areas for $\phi < \phi_{\text{rcp}}$ and to give rise to a loose gel-like network of bubbles connected by contact films  at $\phi > \phi_{\text{rcp}}$, observable only in the absence of gravity. 
The sharp transition to Ostwald ripening behavior at  $\phi\approx\phi^*$ and the prefactor of the coarsening law in this latter regime which is much larger than predicted by most theories call for further investigations. \par
Experiments with systems with   adhesive forces larger than in our present samples are planned in future ISS experiments to take place in 2023. Numerical modeling is also under way to account for the new observed features.

\section{Appendix}
\begin{figure}[h]
    \centering
    \includegraphics[width = 8.5cm]{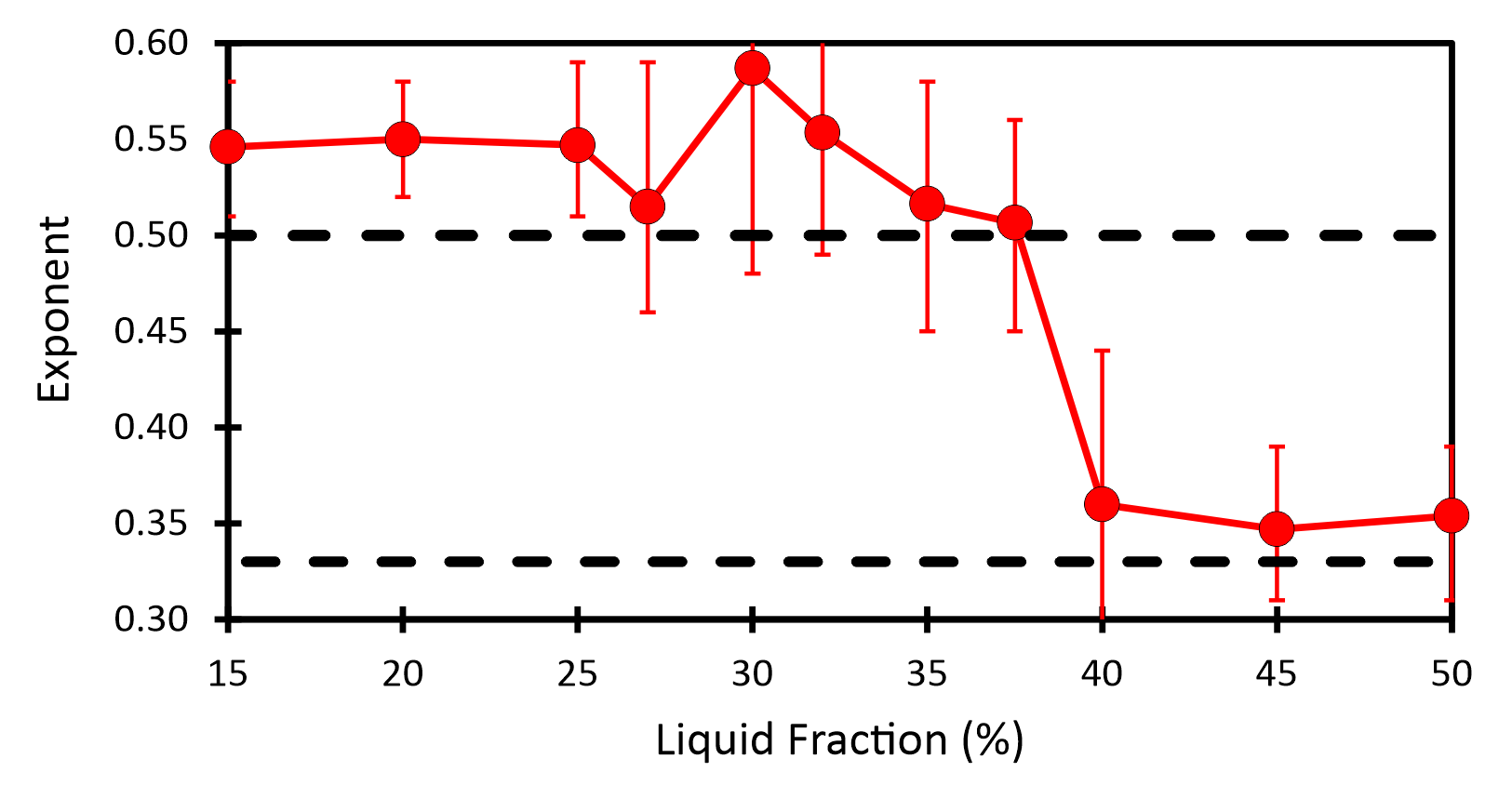}
    \caption{Figure showing the evolution as a function of liquid fraction of the exponent obtained by fitting power law $R_{32} = at^{\alpha}$ on data in the scaling state.
    }
    \label{fig:exponents}
\end{figure}

Fig.~\ref{fig:exponents} illustrates the liquid fraction dependence of the coarsening exponents.

\par
\begin{figure}[h]
    \centering
    \includegraphics[width = 8.5cm]{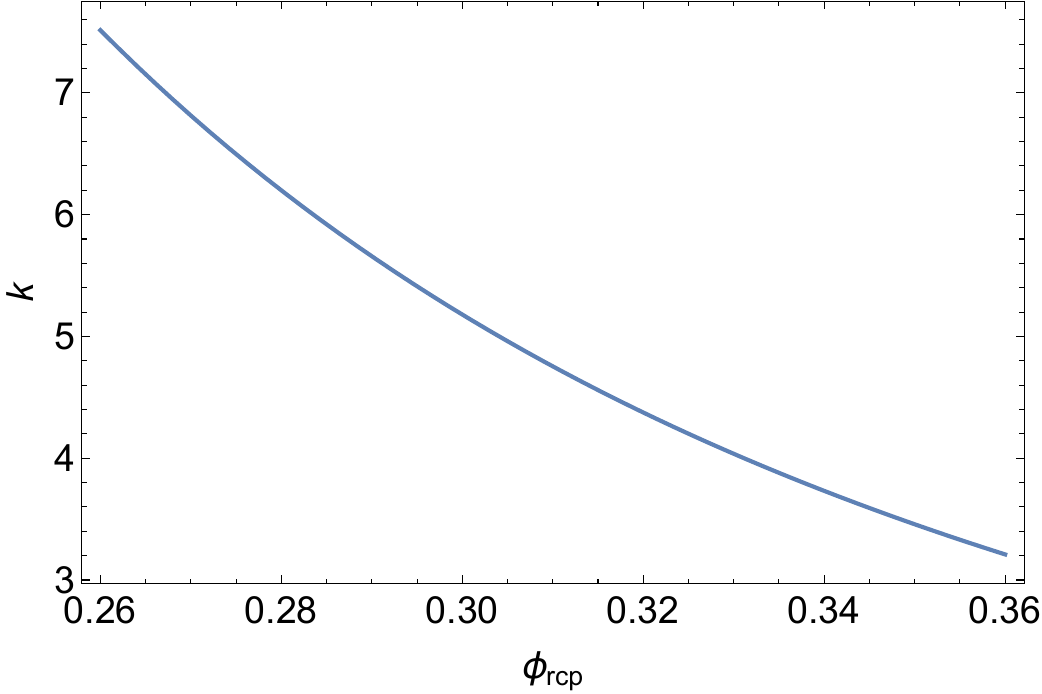}
    \caption{ Coefficient $k$ defined in Eq.~\ref{eq:Pi} versus the jamming liquid fraction $\phi_{\text{rcp}}$.}
    \label{fig:kphi}
\end{figure}

  Fig.~\ref{fig:kphi} illustrates the dependency of the coefficient $k$ defined in Eq.~\ref{eq:Pi} on the jamming liquid fraction $\phi_{\text{rcp}}$, predicted by a calculation described in section~\ref{sec:DiscussionFoamRegime}. Combined with Eq.~\ref{eq:k}, this result enables the prediction of the osmotic pressure of polydisperse foams and emulsions, a result which may be useful well beyond the scope of our present study.  

\section*{Author Contributions}
{M.Pasquet, E.Rio, A.Salonen, D.Langevin, S.Cohen Addad, R.Höhler and O.Pitois contributed to the preparation and follow-up of the ISS experiments. M.Pasquet, N.Galvani and O.Pitois performed the image analysis. M. Pasquet, N.Galvani, O.Pitois and S.Cohen-Addad performed the calculations of the coarsening rates using the different models. S.Cohen-Addad performed the film thickness calculations, R.Höhler performed the ground experiments and the theoretical modelling. All the authors participated in the discussions and writing of the manuscript.}

\section*{Conflicts of interest}
There are no conflicts to declare.

\section*{Acknowledgements}
This work was supported by ESA and CNES, focused on the Soft Matter Dynamics instrument and the space mission Foam-C. Marina Pasquet, Nicolo Galvani and Alice Requier benefited from CNES and ESA PhD grants. The authors are grateful to the BUSOC team for their invaluable help during the ISS experiments. We also want to warmly thank Marco Braibanti and Sébastien Vincent-Bonnieu from ESA, Christophe Delaroche from CNES and Olaf Schoele-Schulz from Airbus for their continuing support. 



\balance

\bibliography{rsc} 
\bibliographystyle{rsc} 


\end{document}